\documentclass[lettersize,journal]{IEEEtran}
\IEEEoverridecommandlockouts

\usepackage{graphicx}
\usepackage{caption}
\usepackage{subcaption}
\usepackage{algorithmic}
\usepackage{amsmath}    	            

\usepackage{cite}            
\usepackage{newtxtext}        
\usepackage{textcomp}   	    
\usepackage[euler]{textgreek}
\usepackage{soul, color}      
\usepackage[svgnames]{xcolor} 
\usepackage[utf8]{inputenc}   
\usepackage[T1]{fontenc}      
\usepackage[shortlabels]{enumitem}  		
\usepackage{url}
\usepackage{microtype}        
\usepackage{tcolorbox}        
\usepackage{marginnote}       
\usepackage{accents}          

\usepackage{amssymb}		            
\usepackage{amsfonts}		            
\usepackage{amsthm}    	                
\usepackage{mathtools}		            
\usepackage{cases}
\usepackage{newtxmath} 	                
\usepackage{bm}                         
\usepackage[ruled,vlined,linesnumbered]{algorithm2e}  
\usepackage{cancel}

\newcommand{\rc}{\textcolor{blue}}

\newcommand{\nonl}{\renewcommand{\nl}{\let\nl\oldnl}}

\newtheorem{remark}{Remark}
\newtheorem{lemma}{Lemma}
\newtheorem{theorem}{Theorem}[section]

\usepackage{graphicx}		
\usepackage{caption}        
\usepackage{subcaption}     
\usepackage{booktabs}   	
\usepackage{siunitx}        
\usepackage[export]{adjustbox}  
\usepackage{mdframed}		

\newcommand{\hX}{\widehat{\C{X}}}
\newcommand{\T}{^{\mathsf{T}}}

\newcommand{\Ptot}{P_\text{tot}}
\newcommand{\B}[1]{\if#1\relax\bm{#1}\else\mathbf{#1}\fi} 
\newcommand{\R}[1]{\mathrm{#1}}						      
\newcommand{\C}[1]{\mathcal{#1}}
\newcommand{\BB}[1]{\mathbb{#1}}

\allowdisplaybreaks 
\title{Consensus-based Distributed Intentional Controlled Islanding of Power Grids}
\author{Francesco Lo Iudice, Ricardo Cardona-Rivera, Antonio Grotta, Marco Coraggio, Mario di Bernardo$^\dagger$
\thanks{%
$^\dagger$Corresponding author; e-mail: mario.dibernardo@unina.it. 
\newline Francesco Lo Iudice, Ricardo Cardona-Rivera, and Mario di Bernardo are with the Department of Electrical Engineering and Information Technology, University of Naples Federico II, 80125, Naples, Italy. Antonio Grotta, Marco Coraggio and Mario di Bernardo are (also) with the Scuola Superiore Meridionale, School for Advanced Studies, Naples, Italy.}}
\date{February 2021}
\begin{document}
\maketitle
\begin{abstract}
The problem of partitioning a power grid into a set of islands can be a solution to restore power dispatchment in sections of a grid affected by an extreme failure. Current solutions to this problem usually involve finding the partition of the grid into islands that minimizes the sum of their absolute power imbalances. This combinatorial problem is often solved through heuristic offline methods. In this paper, we propose instead a distributed online algorithm through which nodes can migrate among islands, self-organizing the network into a suitable partition. 
We prove that, under a set of appropriate assumptions, the proposed solution yields a partition whose absolute power imbalance falls within a given bound of the optimal solution. We validate our analytical results by testing our partitioning strategy on the IEEE 118 and 300 benchmark problems.
\end{abstract}

\section{Introduction}

The penetration of renewable and distributed generation, e.g., \cite{dorfler2015breaking,bidram2012hierarchical,frasca2015distributed}, and the possible occurrence of cascading failures \cite{pourbeik2006anatomy} have made the problem of devising control strategies to govern the operation of power grids of crucial concern.
Examples of problems of interest include those reported in \cite{rocabert2012control,tayyebi2020frequency,arghir2018grid}, \cite{milano2018foundations,dorfler2019distributed} and  \cite{lalor2005frequency,bevrani2010renewable,ulbig2014impact}.

When the control architecture fails to guarantee reliable operation of the transmission grid, last resort strategies have been devised so as to ensure power dispatchment across at least some of its sections.
\emph{Intentional Controlled Islanding} (ICI) strategies address this issue \cite{cris2021Controlled,pahwa2013optimal,sun2003splitting,adibi2006power,fernandez2018intentional} by identifying sections of the grid (or \emph{islands}) that can isolate and operate independently from the rest of the network. Recently, intentional islanding has also been proposed in the framework of distribution networks, see \cite{AhRe:2020review} and references therein, as the presence of storage devices and renewable energy generation allows these grids to be partitioned into {\em networks of microgrids}, e.g., \cite{haddadian2017multi,wang2014coordinated,arefifar2014dg}.

The problem of partitioning a grid into a set of islands  is usually mathematized as a combinatorial problem---see for example \cite{haddadian2017multi,arefifar2013optimum,hasanvand2017new,mohammadi2014scenario}---and sometimes it is recast as a graph optimization problem \cite{fernandez2018intentional,sun2003splitting,adibi2006power}.
Often, solving these problems numerically is cumbersome or inefficient, so that heuristic strategies are frequently used to seek a suboptimal solution, while meeting the required computational time that allows the network to stabilize after a contingency \cite{cris2021Controlled, liu2018controlled,kyriacou2017controlled, wang2010novel}.

In this paper, we propose a different distributed approach to solve the ICI problem, where network nodes can migrate from an island to another so as to \emph{self-organise} into a partition minimizing the power imbalance between different islands and avoiding large amounts of load shedding.
Specifically, starting from some initial partition of the grid, we endow the nodes with the ability of locally estimating the power imbalance of their island and of those neighboring it, so as to decide whether to migrate or not to a different island from their own.

The estimation strategy is completely distributed and decentralized and relies on nodes running a virtual consensus dynamics parameterized so that the consensus equilibrium the nodes reach is proportional to the power imbalance of the island of interest. 
Under suitable assumptions, we analytically show that our migration strategy generates a sequence of partitions that converge in finite time to a configuration whose average absolute power imbalance falls within a certain bound of the minimal one. We validate our strategy by partitioning the IEEE 118 and IEEE 300 test systems, comparing the viable partitions we obtain to others suggested in previous papers in the Literature.

\section{Preliminaries and problem statement}
\label{sec:section_1}

\subsubsection*{Notation}

Given a set $\C{Q}$, we denote by $\left\lvert \C{Q} \right\rvert$ its cardinality; $\B{1}$ is the column vector of ones, with appropriate dimension.

\subsubsection*{Power grid}

We model a power grid as an \emph{undirected connected} graph $\C{G} = (\C{V},\C{E})$, where $\C{V}$ is the set of $n \in \BB{N}_{>0}$ grid nodes (loads or generators) and $\C{E}$ is the set of edges representing transmission lines interconnecting them. 
Without loss of generality, the $n_\R{g} \in \BB{N}_{>0}$ generators are labeled as nodes $1, \dots, n_\R{g}$, while the $n_\R{l} \in \BB{N}_{>0}$ loads as nodes $n_\R{g} + 1, \dots, n$. 
We let $p_i \in \BB{R}$ be the active power generated or consumed at node $i$; $p_i > 0$ if $i$ is a generator, while $p_i\leq 0$ if $i$ is a load.
We let $A$ be the (symmetric) adjacency matrix associated to the graph $\C{G}$; its $(i, j)$-th element $a_{ij}$ being $1$ if $\{i, j\}\in \C{E}$ or $0$ otherwise.

\subsubsection*{Islands and neighbours}

We define an \emph{island} as a connected subgraph $\C{M}_l=(\C{V}_l, \C{E}_l)$ of $\C{G}$, where $\C{V}_l\subseteq\C{V}$ and $\C{E}_l=\left(\C{V}_l\times\C{V}_l\right) \cap \C{E}$. 
Given a set of nodes $\C{V}_l$, we denote by $\C{N}(\C{V}_l)$ the set of \emph{neighbours} of the nodes in $\C{V}_l$, i.e., $\C{N}(\C{V}_l) \coloneqq\lbrace i\in \C{V} \setminus \C{V}_l \mid \exists j \in \C{V}_l:\{i, j\}\in \C{E} \rbrace$. 
We say that island $\C{M}_m$ is a \emph{neighbor} of island $\C{M}_l$ if and only if $\C{N}(\C{V}_m)\cap \C{V}_l \neq \varnothing$. 
Moreover, we denote by $\C{N}_i$ the set of neighbours of node $i$.

\subsubsection*{Grid partitions}

We say that the grid is {\em partitioned} into $n_{\mu} \in \BB{N}_{> 0}$ islands, described by the subgraphs $\C{M}_l, \dots,\C{M}_{n_\mu}$, with corresponding node sets $\C{V}_1,\dots,\C{V}_{n_\mu}$, if $\Pi=\{\C{V}_1,\dots,\C{V}_{n_\mu}\}$ is a \emph{partition} of $\C{V}$.
Additionally, a node, say $i$, in an island, say $\C{M}_l$, is a \emph{boundary node} if $\C{N}_i\cap (\C{V}\setminus \C{V}_l)\neq \varnothing$.
Furthermore, we define the condensed graph $\C{G}^{\Pi} = (\C{V}^{\Pi},\C{E}^{\Pi})$ induced by the partition $\Pi$, where node $l$ in $\C{V}^{\Pi}$ is associated to $\C{V}_l$ in $\Pi$, and an edge $\{l, m\}$ exists in $\C{E}^{\Pi}$ if and only if $\C{V}_l \cap \C{N}(\C{V}_m) \neq \varnothing$.

\subsubsection*{Power imbalance}

The \emph{power imbalance} of an island $\C{M}_l$ is
\begin{equation}\label{eq:P_i}
    P_l \coloneqq \sum_{i\in\C{V}_l}p_i;
\end{equation}
the overall \emph{grid's power imbalance} is \begin{equation}\label{eq:P_tot}
    \Ptot \coloneqq \sum_{i=1}^n p_i = \sum_{l=1}^{n_{\mu}} P_l.
\end{equation}
The power imbalance in \eqref{eq:P_i} is associated to the synchronous frequency deviation of the island from its nominal value, which in turn is related to the frequency's stability \cite{kundur2004definition,dorfler2019distributed}.
Indeed, if the generated power exceeds loads' demand, the frequency increases, and vice-versa.
Excessively large variations in the operating frequency with respect to the nominal one can cause faults. 

\subsubsection*{Control problem}

The problem we study is to find a partition of the power grid $\C{G}$ into $n_\mu\geq 2$ microgrids so as to minimize the \emph{average absolute power imbalance}, defined as
\begin{equation}\label{eq:cost}
    J \coloneqq \frac{1}{n_\mu}\sum_{l=1}^{n_\mu}|P_l|.
\end{equation}
Note that, as $\sum_{l=1}^{n_\mu}|P_l|\geq \left| \sum_{l=1}^{n_\mu}P_l \right| = |\Ptot|$, then
\begin{equation}\label{eq:cost_bound}
    J\geq J^*\coloneqq \left| \frac{\Ptot}{n_{\mu}} \right|.
\end{equation}
The cost function in \eqref{eq:cost} has been used in previous work in the literature on grid partitioning, e.g. \cite{haddadian2017multi,liu2018controlled,cris2021Controlled}, as an indicator of the ability of a power system to satisfy the loads' demand, which is also known as \emph{adequacy} \cite{ArMo:12}.

\section{A consensus based partitioning strategy}

We propose a strategy that, given an initial partition $\Pi(0)$ of the power grid into $n_{\mu}$ islands, uses a consensus algorithm to let the nodes self-organise into a new partition that minimizes $J$, as defined in \eqref{eq:cost}.
In particular, at each step $k$ of the algorithm, one node can migrate between islands. 
We denote by $\Pi(k)$ the partition after $k$ migrations have occurred; $\C{M}_l(k) = (\C{V}_l(k),\C{E}_l(k))$, $l \in \{1,\dots,n_{\mu}\}$ being the corresponding islands, $P_l(k)$, $l \in \{1,\dots,n_{\mu}\}$ their power imbalances, and $J(k)$ the corresponding value of the cost function.

Our strategy is based on two fundamental ingredients:
\begin{enumerate}
\item a \emph{distributed dynamic estimator} based on average consensus dynamics that nodes can use to estimate the power imbalance in their island and in those of their neighbors;
\item a \emph{migration condition} according to which a boundary node can decide whether to migrate from its island to a neighboring one.
\end{enumerate}
Next, we describe the two elements above.

\subsection{Distributed power imbalance estimation}
\label{sec:estimation_strategy}

At any step $k$, each node, say $i$, can obtain an estimate of the power imbalance, say $P_l(k)$, of the island it belongs to or of an island neighboring it, say $\C{M}_l(k) = (\C{V}_l(k),\C{E}_l(k))$, by running a consensus based estimation strategy.

Specifically, let us define the auxiliary graph $\widehat {\C{M}}_{l}(k)\coloneqq ( \widehat{\C{V}}_{l}(k),\widehat{\C{E}}_{l}(k) )$ with
\begin{equation}\label{eq:Aux_nodeset}
    \widehat{\C{V}}_{l}(k) \coloneqq 
    \begin{dcases}
        \C{V}_l(k) \setminus i, & \text{if } i \in {\C V}_l(k), \\
        \C{V}_l(k)\cup i, & \text{if } i \notin {\C V}_l(k),
    \end{dcases}         
\end{equation}
and $\widehat{\C{E}}_{l}(k)\coloneqq ( \widehat{\C{V}}_{l}(k)\times \widehat{\C{V}}_{l}(k) ) \cap \C{E}$. To estimate $P_l(k)$, node $i$ must trigger
the distributed solution of the two \textit{virtual} continuous-time consensus dynamics given by

\begin{subequations}\label{eq:kur_dyn}
\begin{align}\label{eq:kur_dyn_a}
    \dot x_{h}(t) &= p_{h} + \sum_{\{j,h\}\in \C{E}_{l}(k)}  (x_j(t) - x_{h}(t)), 
    \quad \forall {h} \in \C{V}_{l}(k),\\
\label{eq:kur_dyn_b}
    \dot{\widehat{x}}_{h}(t) &= p_{h} + \sum_{\{j,h\}\in \widehat{\C{E}}_{l}(k)}  (\widehat x_{j}(t) - \widehat x_{h}(t)),
    \quad \forall h \in \widehat{\C{V}}_{l}(k),
\end{align}
\end{subequations}
starting from null initial conditions. 
Here, $x_h(t)$ and $\widehat x_{h}(t)$ are the virtual states associated to each node $h \in \C V_{l}(k)$ and $h \in \widehat{\C V}_{l}(k)$, respectively.
\begin{remark}
To run the consensus dynamics \eqref{eq:kur_dyn} in a distributed manner, we assume the virtual states $x_h$ and $\widehat{x}_{h}$ are broadcast to all neighboring nodes $\C{N}_h \cap \C{V}_{l}(k)$. 
\end{remark}
Now, dynamics \eqref{eq:kur_dyn_a} can be recast in matrix form as 
\begin{equation}\label{eq:mat_kur_dyn_a}
    \dot{\pmb{x}}(t) = \B{p} -L\pmb{x}(t), 
\end{equation}
where $\pmb{x}$ is the stack vector of the virtual states $x_h$, $\B{p}$ is the stack vector of the power values $p_h$, and $L$ is the (symmetric) Laplacian matrix associated to $\C{M}_l(k)$.
Let us recall that $\B{1}\T$ is an eigenvector of the symmetric Laplacian $L$, with $0$ as an associated eigenvalue.
To obtain the asymptotic behaviour of \eqref{eq:mat_kur_dyn_a}, we premultiply \eqref{eq:mat_kur_dyn_a} by $\B{1}\T$, obtaining, for all time $t$, 
\begin{equation}\label{eq:first_ord_ss}
    \B{1}\T \dot{\pmb{x}}(t) = \B{1}\T \B{p} = P_l(k).
\end{equation}
Moreover, differentiating \eqref{eq:mat_kur_dyn_a}, we obtain the dynamical system $\ddot{\pmb{x}}(t) = - L \dot{\pmb{x}}(t)$,
whose dynamics, determined by the spectral properties of $L$, are such that 
\begin{equation}\label{eq:sec_ord_ss}
    \lim_{t \rightarrow \infty} \dot{\pmb{x}}(t) \in \R{span}(\B{1}).
\end{equation} 
Altogether, \eqref{eq:first_ord_ss} and \eqref{eq:sec_ord_ss} imply that
$
\lim_{t \rightarrow \infty} \dot{\pmb{x}}(t) = \B{1} \omega_l
$,
where
\begin{equation}\label{eq:ss_freq_prel_a}
\omega_l \coloneqq \frac{P_l(k)}{|\C{V}_{l}(k)|}, 
\end{equation}
Similarly, from \eqref{eq:kur_dyn_b}, we obtain that $\lim_{t \rightarrow \infty}\dot{\widehat{\pmb{x}}}(t) = \B{1} \widehat{\omega}_l$, with
\begin{equation}\label{eq:ss_freq_prel_b}
\widehat\omega_l \coloneqq \frac{1}{|\widehat{\C{V}}_{l}(k)|} \sum_{j \in \widehat{\C{V}}_{l}(k)}p_j.   
\end{equation} 
Exploting \eqref{eq:Aux_nodeset}, \eqref{eq:ss_freq_prel_b} can be recast as
\begin{equation}\label{eq:ss_freq}
\widehat\omega_l =  \begin{dcases}
          \tfrac{1}{|\C{V}_{l}(k)|-1} \left( P_l(k) - p_i \right), & \text{if } i \in \C{V}_l(k),\\
          \tfrac{1}{|\C{V}_{l}(k)|+1} \left( P_l(k) + p_i \right), & \text{if } 
          i \not\in \C{V}_l(k).
\end{dcases}
\end{equation}
Then, \eqref{eq:ss_freq_prel_a} and \eqref{eq:ss_freq} can be solved together for the unknowns $P_l(k)$ and $|\C{V}_{l}(k)|$, obtaining
\begin{subequations}\label{eq:pow_bal_est}
\begin{equation}\label{eq:pow_bal_est_a}
    P_l(k) = a_l\omega_l\frac{p_i -\widehat{\omega}_l}{\widehat{\omega}_l - \omega_l},
\end{equation}
\begin{equation}\label{eq:pow_bal_est_b}
    |\C{V}_l(k)| =  a_l\frac{p_i -\widehat{\omega}_l}{\widehat{\omega}_l - \omega_l},
\end{equation}
\end{subequations}
with 
\begin{equation*}
    a_l = \begin{dcases}
         -1, & \text{if } i \in \C{V}_l(k),\\
          1, & \text{if } i \not\in \C{V}_l(k).
    \end{dcases}
\end{equation*}

From \eqref{eq:pow_bal_est_a}, to estimate $P_l(k)$, node $i$ needs to compute $\omega_l$ and $\widehat \omega_l$. 
To do so in a distributed fashion, node $i$ starts the distributed computation of the consensus dynamics \eqref{eq:kur_dyn_a} and \eqref{eq:kur_dyn_b} by broadcasting its virtual states $x_i$ and $\widehat{x}_i$ to 
the nodes in ${\C{N}}_i \cap \C{V}_{l}(k)$.
In turn, each of these starts sharing its virtual state with its neighbors (within $\C{V}_{l}(k)$),
until all nodes in $\C{V}_{l}(k)$ join the distributed simulation. 
Note that the aforementioned procedure can be conducted through one-hop communication if each node $h$ has knowledge of the index $l \in \{1,...,n_\mu\}$ of the island it belongs to, and of its consumed or generated power $p_h$. 
Obviously, in a practical implementation, the grid nodes should be equipped with sufficient computational and communication capabilities to run the virtual consensus dynamics on a timescale that is compatible with the grid requirements.

In what follows, we will show how the network nodes can exploit this estimation process to self-organise into a partition of the power network whose power imbalance \eqref{eq:cost} is rendered minimal.

\begin{figure}[t]
    \centering
     \includegraphics[scale=0.42, trim = {6.7cm 1cm 6.5cm 1.6cm},clip]{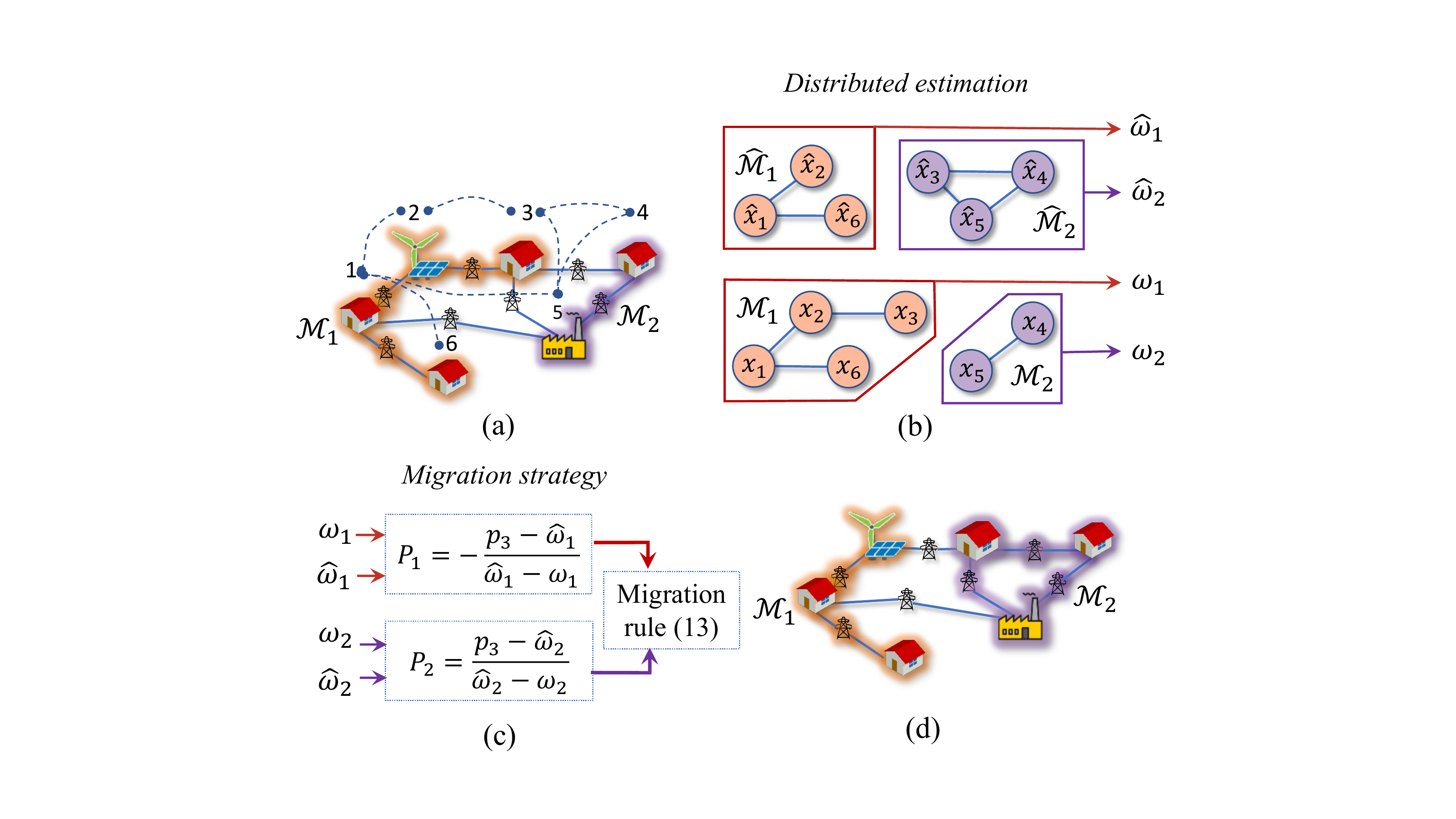}
          \label{fig:cyberlayers_a}
    \caption{(a) Initial partition of the power network, with dashed lines representing the communication links among nodes; the topology being equal to that of the power network itself. 
    (b) Boundary node 3 triggers the distributed simulation of the virtual consensus dynamics  in \eqref{eq:kur_dyn} for both islands $\C M_1 $ and $\C M_2$.
    (c) The migration rule \eqref{eq:migration_rule} is used to decide whether to migrate from island $\C{M}_1$ to $\C{M}_2$ and 
    (d) a new partition is eventually generated.}
    \label{fig:example}
\end{figure}

\subsection{Migration Condition}

A boundary node (see~§ \ref{sec:section_1}), say $i$, in island $\C{M}_m(k)$, can decide whether to migrate to a neighboring island $\C{M}_l(k)$ (see Figure \ref{fig:example}) by assessing the power imbalances $P_l(k)$ and $P_m(k)$ (computed through our estimation strategy in~§ \ref{sec:estimation_strategy}).

Specifically, at step $k$, node $i$ will migrate from $\C{M}_m(k)$ to $\C{M}_l(k)$ if and only if
\begin{subnumcases}{}\label{eq:migration_rule}
    \min (P_l(k),P_m(k)) < \min (P_l(k+1),P_m(k+1)),
    \label{eq:migration_rule1} \\
    \C{M}_m(k+1) \ \R{is \ connected},
    \label{eq:migration_rule2}
\end{subnumcases}
with
\begin{subequations}\label{eq:migration_rule_c}
\begin{align}
    \label{eq:migration_rule_c_1}
    {P}_l(k+1) &= {P}_l(k) + p_i,\\
    \label{eq:migration_rule_c_2}
    {P}_m(k+1) &= {P}_m(k) - p_i,\\
    \C{V}_l(k+1) &= \C{V}_l(k) \cup i,\\
    \C{V}_m(k+1) &= \C{V}_m(k) \setminus \{i\}.
\end{align}    
\end{subequations}

\begin{remark}\label{remark_sync}
Condition \eqref{eq:migration_rule2} concerning connectivity can be ensured using the estimation strategy in~§ \ref{sec:estimation_strategy}.
Indeed, given an island $\C{M}_l(k)$, if there exists a node $i\in\C{V}_l(k)$ such that $\C{M}_l(k) \setminus \{i\}$ is not connected, the virtual derivatives $\dot{\widehat{x}}_{h}$ of its neighbors in $\widehat{\C{M}}_{l}$ (see \eqref{eq:kur_dyn_b}) will in general converge to different values, thus providing a warning signal.
\end{remark}

\subsection{Migration Algorithm} \label{algorithm_description}

According to our decentralised partitioning strategy, starting from some initial partition at step $k = 0$, each boundary node must trigger the distributed estimation of the power imbalance of the island it belongs to and of its neighboring islands by running the virtual consensus dynamics \eqref{eq:kur_dyn}.
Then, depending on these power imbalances, exploiting the migration condition \eqref{eq:migration_rule}, the boundary nodes will decide whether to migrate or not to neighboring islands.

For the sake of clarity, we illustrate the process by referring to the exemplary situation depicted in Figure \ref{fig:example}, where a grid with $n = 6$ nodes is initially partitioned in $n_\mu = 2$ islands,  $\mathcal{M}_1(0)$ and $\mathcal{M}_2(0)$ (Figure \ref{fig:example}(a)).
Then, each boundary node, as for instance node $3 \in \C{V}_1(0)$, must decide whether to migrate to the other island ($\C{M}_2$) or not. 
To this aim, node 3 triggers the distributed estimation of the power imbalances $P_1(0)$ and $P_2(0)$ in both the islands $\mathcal{M}_1(0)$ and $\mathcal{M}_2(0)$ (see Figure \ref{fig:example}(b)), by running two virtual consensus processes of the form \eqref{eq:kur_dyn} involving all nodes belonging to each of the islands. 
Once a steady state in the distributed simulation of \eqref{eq:kur_dyn} has been reached, node 3 uses the pairs $(\omega_1,\hat \omega_1)$ and $(\omega_2,\hat \omega_2)$ to estimate $P_1(0)$ and $P_2(0)$, which it then uses to evaluate the migration condition \eqref{eq:migration_rule} (Figure \ref{fig:example}(c)) to assess whether to migrate from $\mathcal{M}_1$ to $\mathcal{M}_2$. 
Once this decision is taken, a new partition is generated (Figure \ref{fig:example}(d)).

In general, our strategy prescribes that the grid nodes get involved in all the (possibly multiple) distributed consensus processes invoked according to \eqref{eq:kur_dyn} by the boundary nodes of their island or of neighboring ones so as to allow the estimation of the power imbalances of interest. 
Hence, at any time, each node will have a number of virtual states corresponding to the number of estimation processes it is asked to contribute to. These steps are summarized in Algorithm \ref{alg:Migr_alg_a}. 
Additionally, as soon as a node becomes a boundary node (see~§ \ref{sec:section_1}), it must trigger additional virtual dynamics to decide whether to migrate or not from its island to a neighboring one.
This additional procedure is summarized in Algorithm \ref{alg:Migr_alg_b}.

The following Lemma and Theorem, whose proofs are given in Section \ref{sec:proof}, state that the migration process governed by rule \eqref{eq:migration_rule} generates a finite sequence $\{\Pi(k)\}_{k \in \{0, \dots, K\}}$ of $K \in \BB{N}$ migration steps, and give a bound on the difference between the cost  $J(K)$ of the final partition and the optimal cost $J^*$ computed in \eqref{eq:cost_bound}.

\begin{lemma}\label{lemma1}
If 
    \begin{equation}\label{eq:neigh_bound}
        |P_l(k)-P_m(k)|\leq \bar{p} \quad  \forall l,m:\C{N}(\C{V}_m(k))\cap \C{V}_l(k) \neq \varnothing,
    \end{equation}
    where $\bar{p} \coloneqq \max_{i \in \C{V}}|p_i|$, then 
    \begin{equation}\label{eq:fin_bound}
        J(k)-J^* \leq
           \frac{2}{n_{\mu}}\left( \sum_{l = l^* + 1}^{n_\mu}  p^* + \bar p \left( l-\frac{n_{\mu}+1}{2}\right)\right)-(p^*+|p^*|),
    \end{equation}
with
\begin{equation}\label{eq:l_star}
l^*=
       \left\lceil -\frac{p^*}{\bar p} + \frac{n_{\mu}+1}{2}\right\rceil,
\end{equation}
and $p^*\coloneqq\Ptot/n_\mu$.
\end{lemma}

\begin{theorem}\label{thm:main}
    Assume that at each step $k$ there exist a node $i$ and islands $\C{M}_l(k)$ and $\C{M}_m(k)$ (that is a triplet $( l,m,i )$) such that
    \begin{subequations}\label{eq:hypothesis}
    \begin{equation}\label{eq:hypothesis_1}
        \left \{ \begin{array}{c}
          i \in \left\{\C{V}_m(k)\cap\C{N}(\C{V}_l(k))\right\}  \\
         \wedge \\ 
         \C{M}_m(k)\setminus i \text{ is connected }  
          \end{array} \right.
    \end{equation} 
    and 
    \begin{equation}\label{eq:hypothesis_2_1}
         \left \{ \begin{array}{c}
         P_l(k)>P_m(k) \; \wedge \; p_i<0  \\
         \lor \\ 
         P_l(k)<P_m(k) \; \wedge \; p_i>0.
        \end{array} \right.
   \end{equation}
   \end{subequations}
    Then, the sequence $\Pi(k)$ obtained under the migration rule \eqref{eq:migration_rule} is finite and converges in $K < + \infty$ steps to a partition $\Pi(K)$ such that $J(k)$ fulfills \eqref{eq:fin_bound} at $k=K$.
    \end{theorem}

In the following section, we validate the strategy numerically. 
A formal proof of convergence is provided later in Section \ref{sec:proof}.
\makeatletter
\patchcmd{\@algocf@start}
  {-1.5em}
  {0pt}
  {}{}
\makeatother

\begin{algorithm}[t]
Broadcast all virtual states to neighboring nodes

Obtain virtual states from neighboring nodes

Integrate \eqref{eq:kur_dyn} for all simulations where $h$ is involved
\caption{Default routine for any node $h$.}
\label{alg:Migr_alg_a}
\end{algorithm}

\begin{algorithm}[t]
Communicate with the nodes in $\mathcal{N}_h\cap\mathcal{V}_m$ to trigger a distributed simulation of \eqref{eq:kur_dyn}

\For{ $l:\C{N}_h\cap \C{V}_l\neq \varnothing$}{
Communicate with the nodes in $\mathcal{N}_h\cap\mathcal{V}_l$ to trigger a distributed simulation of \eqref{eq:kur_dyn}

Wait for steady state in such simulations

Estimate $P_m$ and $P_l, \ \forall l : \C{N}_h\cap \C{V}_l\neq \varnothing$ using \eqref{eq:pow_bal_est}

Decide whether to migrate from $\C{M}_m$ to $\C{M}_l$ via \eqref{eq:migration_rule} 
}
\caption{Additional steps for a \emph{boundary node} $h \in \C{V}_m$.}

\label{alg:Migr_alg_b}
\end{algorithm}

\section{Numerical Validation}

We demonstrate the effectiveness of our algorithm by deploying it to partition the IEEE 118 and 300 testbed cases \cite{Test_Cases}.
The nodal power values $p_i$ are computed by solving an Optimal Power Flow (OPF) problem, leveraging \textsc{Matpower} 6.0 \cite{zimmerman2010matpower}. 
As the test cases include nodes with null nodal power $p_i=0$, we allow for these nodes to migrate from their island, say $\C{M}_m(k)$, to a neighboring island, say $\C{M}_l(k)$, as long as (i) their migration does not render $\C{M}_m(k)$ disconnected and (ii) $P_l(k) \neq P_l(k'), \forall k' < k : i\in\C{V}_l(k')$.

To apply our partitioning strategy (Algorithms \ref{alg:Migr_alg_a}, \ref{alg:Migr_alg_b}), we need some initial partitions $\Pi(0)$, and to test our algorithm under different conditions, we considered multiple possible $\Pi(0)$. 
In some cases, we took as $\Pi(0)$ some selected partitions from \cite{kyriacou2017controlled,bialek2021tree,cris2021Controlled}. 
In other cases, we used what we call the SSRP+BFS approach to generete $\Pi(0)$.
Namely, we first employ the Search Space Reduction Procedure \cite{kyriacou2017controlled}, which generates a spanning tree connecting groups of coherent generators (these are taken from \cite{kyriacou2017controlled}). 
Then, the remaining nodes are aggregated to the tree using the Breadth-First Search algorithm \cite{cormen2009introduction}. 

\begin{remark}
Throughout our numerical analysis, whenever a node, say $i \in \C{V}_m(k)$, can choose to migrate to more than one island, it will select the one maximizing the difference
\[ \Delta P_{l} =\min \{P_l(k) + P_i, P_m(k) - P_i\} - \min \{P_l(k), P_m(k)\}.\]
This choice ensures that the average absolute power imbalance is improved the most after the migration.
\end{remark}

\setlength{\tabcolsep}{4pt} 
\begin{table*}[t]
\begin{center}
\resizebox{1\textwidth}{!}{%
\begin{tabular}{@{}llllllllllll@{}}
\toprule
Case       & $n_\mu$ & $K$ & Cut-set at $\Pi(0)$ & $\Pi(0)$ & Cut-set at $\Pi(K)$  & $J(0)$ & $J(K)$ & $J^*$ & $P_l(0)$ & $P_l(K)$ & Bound \eqref{eq:fin_bound} \\ 
\midrule
IEEE 118   &  2 & 10 & $\begin{array}{c}\{24\text{-}70, 34\text{-}43, \\37\text{-}40, 38\text{-}65,\\ 39\text{-}40, 71\text{-}72\}\end{array}$ & SSRP+BFS& $\begin{array}{c}\{15\text{-}19, 18\text{-}19,\\ 19\text{-}34, 23\text{-}25,\\ 23\text{-}32, 30\text{-}38,\\ 37\text{-}38, 37\text{-}39,\\ 37\text{-}40, 43\text{-}44\}\end{array}$ &$120.5$ & $58.25$ &$58.25$& $\begin{array}{c}\{-74.26,\\190.75\}\end{array}$ & $\begin{array}{c}\{53.74,\\62.75\}\end{array}$ & 213.14 \\ \hline

IEEE 118   & 2 & 9 & $\begin{array}{c}\{1\text{-}2, 3\text{-}12,\\ 5\text{-}8, 6\text{-}7,\\ 11\text{-}12, 15\text{-}17,\\ 15\text{-}19, 24\text{-}70,\\ 30\text{-}38,34\text{-}36,\\ 44\text{-}45, 70\text{-}71\}\end{array}$ & \cite{cris2021Controlled} &   $\begin{array}{c}\{4\text{-}5, 5\text{-}11,\\ 11\text{-}12, 15\text{-}17,\\ 15\text{-}19, 30\text{-}38,\\ 34\text{-}37, 35\text{-}37,\\ 43\text{-}44, 69\text{-}70,\\ 70\text{-}75, 74\text{-}75 \}\end{array} $ &     $265.5$ & $58.25$ & $58.25$ &$\begin{array}{c}\{-258.25,\\374.74\}\end{array}$  &$\begin{array}{c}\{ 65.75,\\ 50.74\}\end{array}$ & 213.14\\ \hline

IEEE 118   &  3 & 7 &$\begin{array}{c}\{24\text{-}70, 34\text{-}43,\\ 37\text{-}40, 38\text{-}65,\\ 39\text{-}40, 68\text{-}81,\\ 69\text{-}77, 71\text{-}72,\\ 75\text{-}77, 76\text{-}118\}\end{array}$ &   SSRP+BFS &$\begin{array}{c}\{19\text{-}34, 21\text{-}22,\\ 23\text{-}25, 23\text{-}32,\\ 30\text{-}38, 34\text{-}36,\\ 34\text{-}37, 37\text{-}38,\\ 37\text{-}39, 37\text{-}40,\\ 68\text{-}81, 69\text{-}77\\ 75\text{-}77, 76\text{-}118\}\end{array}$ &$80.34$ & $38.83$ & $38.83$ & $\begin{array}{c}\{ -74.26,\\1.98,\\188.77 \}\end{array}$ & $\begin{array}{c}\{53.74,\\ 1.98,\\ 60.77 \}\end{array}$& 335.97 \\ \hline

IEEE 118   &  3 & 8 & $\begin{array}{c}\{24 \text{-} 70, 24 \text{-} 72,\\ 38 \text{-} 65, 40 \text{-} 42,\\ 41 \text{-} 42, 44 \text{-} 45,\\
69 \text{-} 77, 75 \text{-} 77,\\ 81 \text{-} 80, 118 \text{-} 76\}\end{array}$ & \cite{kyriacou2017controlled}& $\begin{array}{c}\{24\text{-}70, 42\text{-}49,\\ 44\text{-}45, 61\text{-}64,\\ 63\text{-}64, 65\text{-}66,\\ 65\text{-}68, 69\text{-}77,\\ 71\text{-}72, 75\text{-}77,\\ 76\text{-}118, 80\text{-}81\}\end{array}$ & 147 & 38.83 & 38.83 & $\begin{array}{c}\{  -199.26\\    313.77\\    1.98\}\end{array}$ & $\begin{array}{c}\{ 83.66,\\30.86, \\1.98\}\end{array}$& 335.97 \\

\bottomrule
\end{tabular}
}
\caption{Results after applying Algorithms \ref{alg:Migr_alg_a} and \ref{alg:Migr_alg_b} to the IEEE 118 test case, considering different initial partitions $\Pi(0)$. 
Power values are reported in MW. Note that bound \eqref{eq:fin_bound} is computed for $k=K$.}
\label{Tab:migration_results_118}
\end{center}
\end{table*}

\subsection{IEEE 118 bus system}

We used our Algorithm \ref{alg:Migr_alg_a}-\ref{alg:Migr_alg_b}
to partition the IEEE 118 test system in $n_{\mu}=2$ and $n_{\mu}=3$ islands, considering only $n_\R{g}=19$ generators (excluding the reactive compensators).
We assume that the migration process is triggered by a three phase solid ground fault at bus 15 forcing line 14-15 to disconnect.
With $n_{\mu}=2$, we considered as initial partition $\Pi(0)$ the one generated by SSRP+BFS and the final partition reported in \cite{cris2021Controlled};
with $n_{\mu}=3$, we considered as $\Pi(0)$ the partition generated by SSRP+BFS and the final one reported in \cite{kyriacou2017controlled}.
All relevant information and the results are reported in Table \ref{Tab:migration_results_118}.

\begin{figure}[t]
    \begin{subfigure}{0.5\textwidth}
        \centering
        \includegraphics[scale=0.4]{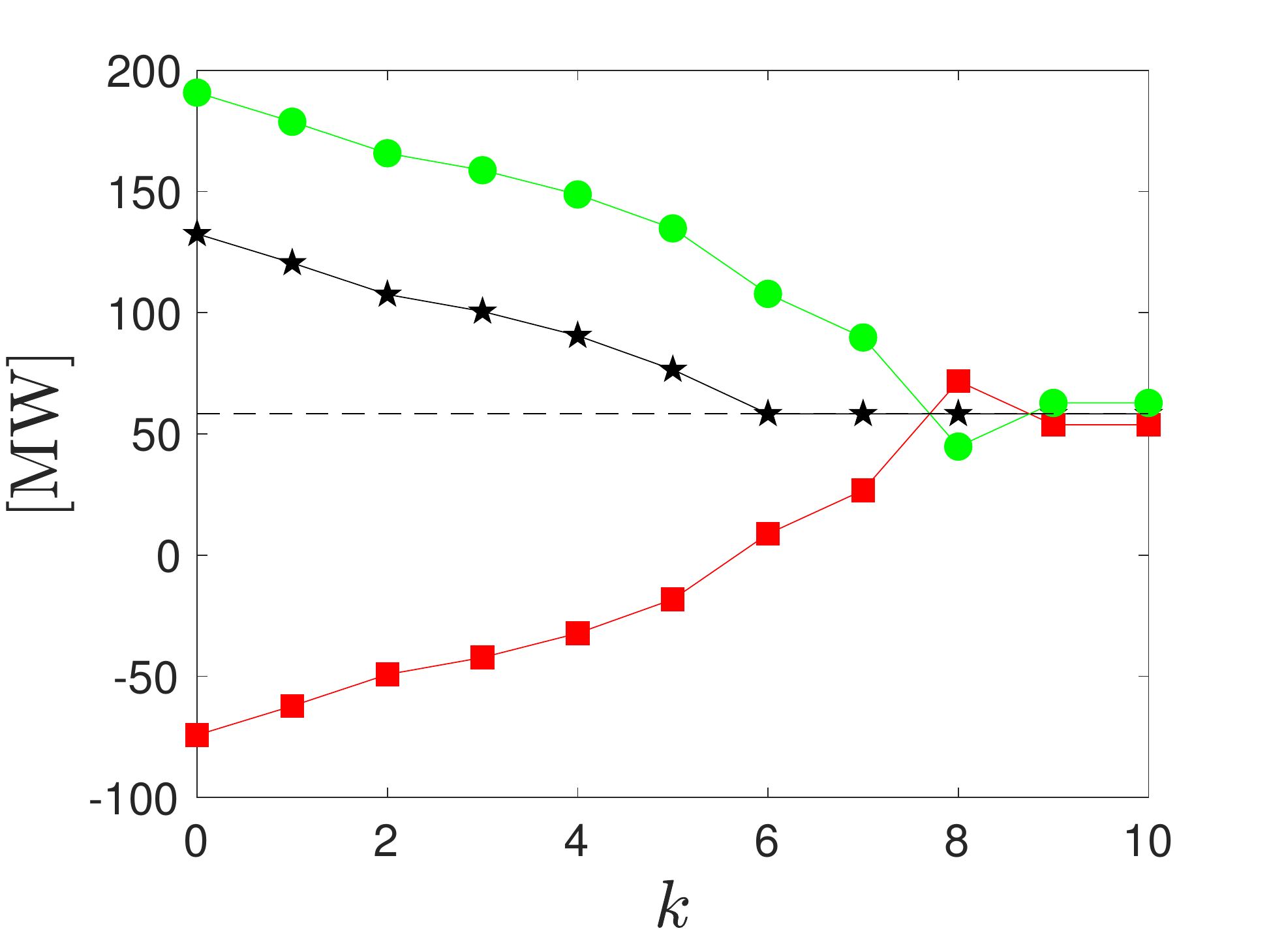}
        \caption{}
        \label{fig:cost_118}
    \end{subfigure}
    \begin{subfigure}{0.5\textwidth}
        \centering
        \includegraphics[scale=0.6, trim = {4cm 1.7cm 3cm 2cm},clip]{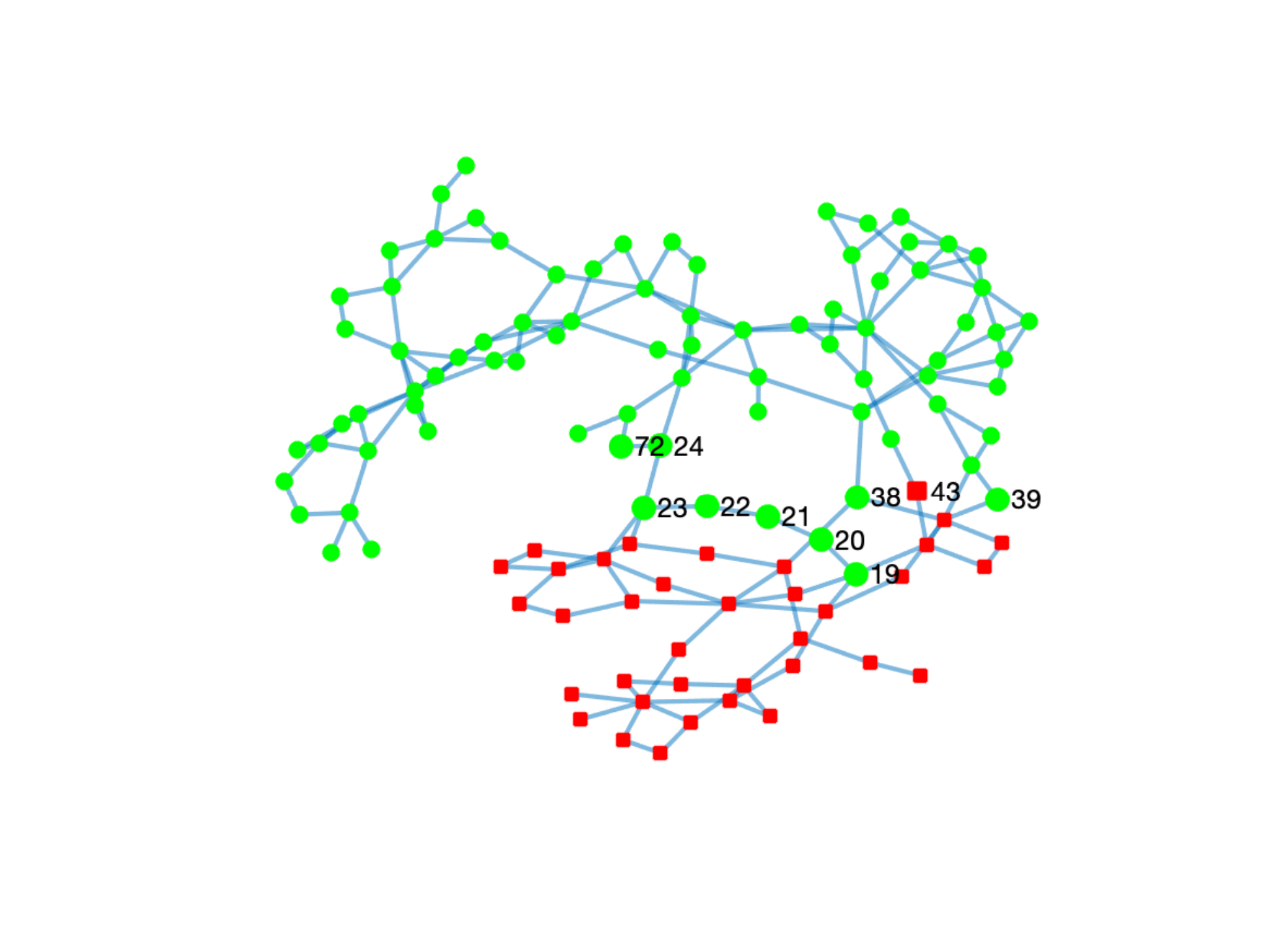}
        \caption{}
        \label{fig:graph_118}
    \end{subfigure}
  \caption{Partitioning of the IEEE 118 test system into $n_\mu=2$ islands, through Algorithms \ref{alg:Migr_alg_a} and \ref{alg:Migr_alg_b}. 
  (a) $P_1(k)$ (red squares), $P_2(k)$ (green circles), $J(k)$ (black stars), and $J^*$ (dashed line); all are in MW.
  (b) Final network partition $\Pi(K)$; red square denote $\C{V}_1(K)$, while green circles denote $\C{V}_2(K)$.
  Nodes $72, 24, 23, 22, 21, 39, 20, 19, 38$ migrated from  $\C{M}_1 $ to $\C{M}_2 $ in the given order, while node $43$ migrated from $\C{M}_2 $ to $\C{M}_1 $ at $k = 9$. Note that the last migration does not change the power imbalances as the it involves node $38$ with nodal power is zero.}
  \label{fig:Partition_118}
\end{figure}

We observe that the proposed algorithm is indeed capable of converging in all cases towards partitions that minimize $J$, as $J(K) = J^*$.
As a representative example, we depict in Figure \ref{fig:Partition_118} the case that $n_{\mu}=2$ and $\Pi(0)$ is generated by SSPR+BFS; namely, Figure \ref{fig:cost_118} portrays the power imbalances $P_1(k)$ and $P_2(k)$ at the various steps, while the final partition $\Pi(K)$ is reported in Figure \ref{fig:graph_118}.
Note that from the OPF results we have $\max_i |p_i|= 542.78$ MW and $J^*=58.25$ MW and thus the bound given in Theorem \ref{thm:main} is satisfied as $|J(K) - J^*| = 0$ (see Table \ref{Tab:migration_results_118}).

\subsection{IEEE 300 bus system}
We used Algorithms \ref{alg:Migr_alg_a} and \ref{alg:Migr_alg_b} to partition the IEEE 300 test system in $n_{\mu}=3$ and $n_{\mu}=4$ islands, assuming a failure affects line 194-195. 
With $n_{\mu} = 3$, as $\Pi(0)$ we consider the SSRP+BFS partition and an arbitrary partition reported in Table \ref{Tab:migration_results_300};
with $n_{\mu} = 4$, as $\Pi(0)$ we consider the SSRP+BFS partition and that from \cite{kyriacou2017controlled}.
In both cases, the groups of coherent generators were selected as in Table II of \cite{kyriacou2017controlled}.
All relevant information and the results are reported in Table \ref{Tab:migration_results_300}.

Again, in all cases, our algorithm is capable of finding an optimal partition, as $J(K) = J^*$; notably, the SSRP+BFS initial partitions and that in \cite{kyriacou2017controlled} are already optimal, but our algorithm is able to further decrease the standard deviation between the power imbalances of the three islands (See Table \ref{Tab:migration_results_300}).

\begin{figure}[t]
    \begin{subfigure}{0.5\textwidth}
        \centering
        \includegraphics[scale=0.4]{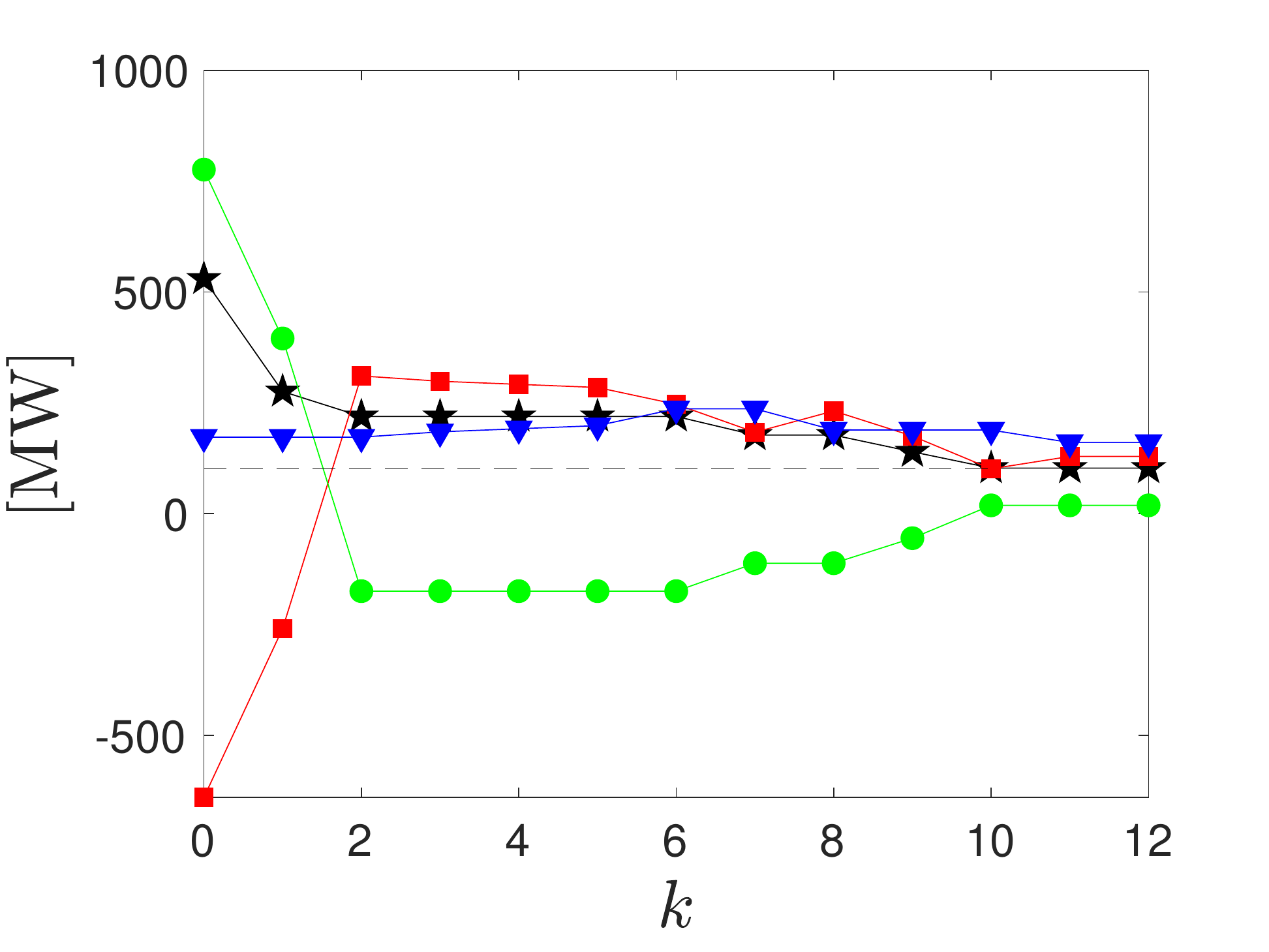}
        \caption{}
        \label{fig:cost_300}
    \end{subfigure}
    
    \begin{subfigure}{0.5\textwidth}
        \centering
        \includegraphics[scale=0.5, trim = {6cm 2cm 4cm 3cm},clip]{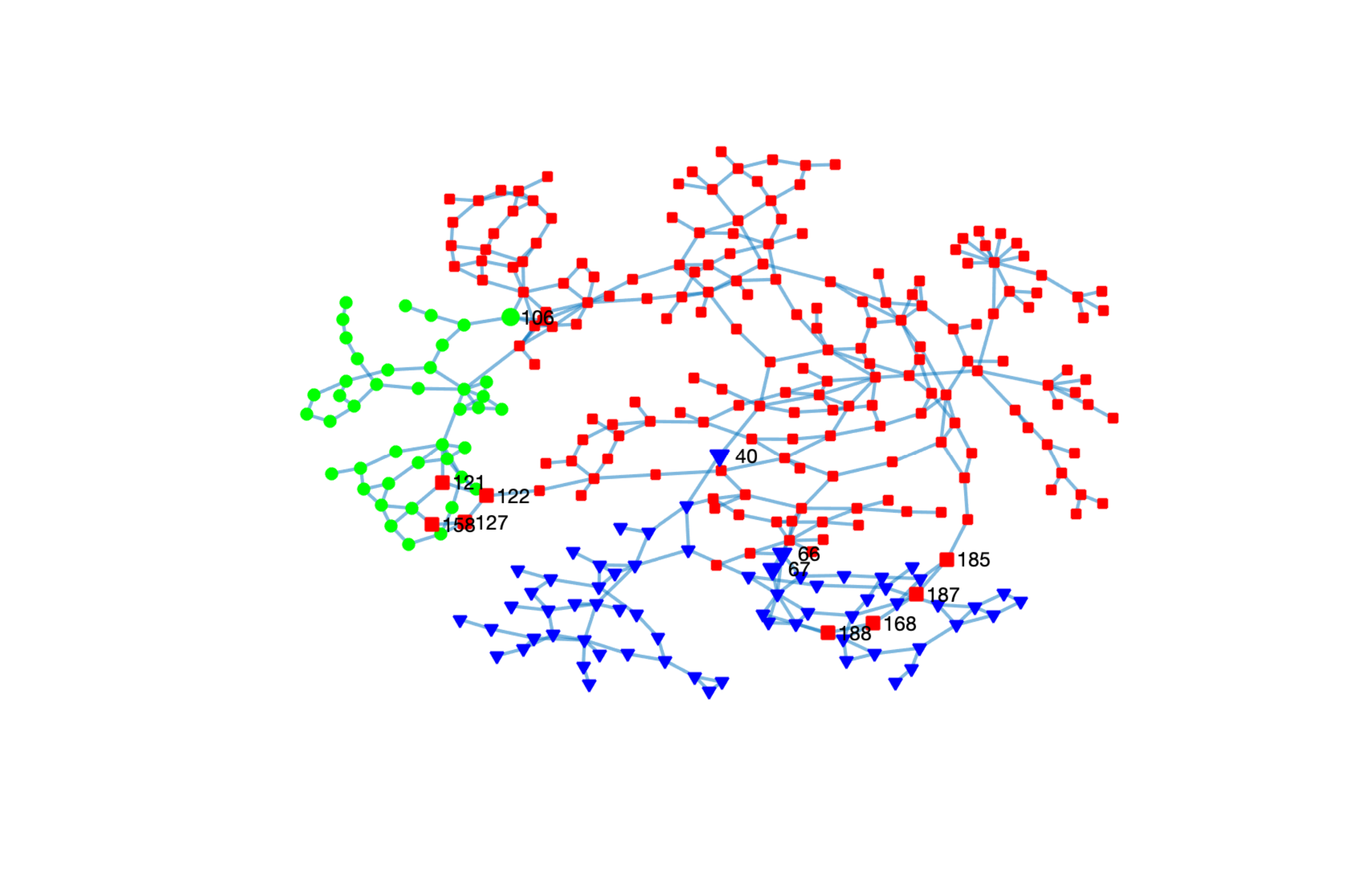}
        \caption{}
        \label{fig:graph_300}
    \end{subfigure}
    
  \caption{Partitioning of the IEEE 300 test system into $n_\mu = 3$ islands, through Algorithms \ref{alg:Migr_alg_a} and \ref{alg:Migr_alg_b}.
  (a) $P_1(k)$ (red squares), $P_2(k)$ (green circles), $P_3(k)$ (blue triangles), $J(k)$ (black stars), and $J^*$ (dashed line).
  (b) Final network partition $\Pi(K)$; red squares denote $\C{V}_1(K)$, green circles denote $\C{V}_2(K)$, and blue triangles denote $\C{V}_3(K)$. The nodes' migration order is $ 106, 122, 185, 187, 168, 188, 127, 66, 121, 158, 67$ and $40$.}
  \label{fig:Partition_300}
\end{figure}

In Figure \ref{fig:Partition_300}, we report the representative case that $n_{\mu}=3$ and $\Pi(0)$ is the arbitrary one. 
The power imbalances $P_1(k)$, $P_2(k)$, $P_3(k)$ are depicted in Figure \ref{fig:cost_300}, while the final partition $\Pi(K)$ is portrayed in Figure \ref{fig:graph_300}.
Interestingly, across all our numerical experiments, not only does our algorithm ensure fulfillment of the bound given in Theorem \ref{thm:main}, but it also always ensures $J(K)=J^*$, and in all cases it succeeds in reducing the standard deviation among the power imbalances of the islands with respect to that of the initial partition (see Table \ref{Tab:migration_results_300}).

Finally, we note that, as shown in Table \ref{Tab:migration_results_300}, for a given test case and a desired number of islands $n_{\mu}$, there are multiple optimal solutions minimizing $J$. 
This opens the possibility of developing a multi-objective partitioning strategy, which might be the subject of future study.

\begin{table*}[t]
\begin{center}
\resizebox{1\textwidth}{!}{%
\begin{tabular}{@{}llllllllllll@{}}
\toprule
Case       & $n_\mu$ & $K$ & Cut-set at $\Pi(0)$ & $\Pi(0)$ & Cut-set at $\Pi(K)$  & $J(0)$ & $J(K)$ & $J^*$ & $P_l(0)$ & $P_l(K)$ & Bound \eqref{eq:fin_bound}  \\
\midrule
IEEE 300  & 3  & 3 &$\begin{array}{c}\{3\text{-}129, 7\text{-}110,\\ 40\text{-}68, 54\text{-}123,\\ 57\text{-}66, 66\text{-}190,\\ 67\text{-}190, 68\text{-}73,\\ 185\text{-}186\}\end{array}$& SSRP+BFS& $\begin{array}{c}\{ 3\text{-}129, 40\text{-}68,\\ 54\text{-}123, 57\text{-}66,\\ 64\text{-}67, 66\text{-}190,\\ 68\text{-}73, 109\text{-}110,\\ 184\text{-}185, 185\text{-}187 \}\end{array}$  &  $102.92 $ & $102.92 $ & $102.92 $ &  $\begin{array}{c}\{6.11,\\129.98,\\ 172.65\}\end{array}$ &  $\begin{array}{c}\{6.11,\\ 145.98,\\156.65\}\end{array}$& 1254.95\\ \hline

IEEE 300  & 3  & 12 &$\begin{array}{c}\{ 40\text{-}68, 57\text{-}66,\\ 66\text{-}190, 67\text{-}190,\\ 68\text{-}73, 106\text{-}113,\\ 112\text{-}116, 122\text{-}123,\\ 185\text{-}186\}\end{array}$ & Arbitrary &$\begin{array}{c}\{ 36\text{-}40, 39\text{-}40,\\ 61\text{-}66, 64\text{-}67,\\ 65\text{-}66, 68\text{-}73,\\ 105\text{-}106, 106\text{-}107,\\ 106\text{-}147, 112\text{-}116,\\ 119\text{-}121, 121\text{-}154,\\ 122\text{-}124, 122\text{-}128,\\ 127\text{-}157, 154\text{-}158,\\ 157\text{-}158, 168\text{-}189,\\ 172\text{-}187, 177\text{-}188,\\ 184\text{-}185\}\end{array}$  &  $529.49 $ & $102.92 $ & $102.92 $ &  $\begin{array}{c}\{-639.87,\\775.96, \\172.65\}\end{array}$ &  $\begin{array}{c}\{129.21,  \\  18.89, \\   160.65\}\end{array}$& 1254.95\\ \hline

IEEE 300  & 4  & 5 &$\begin{array}{c}\{3\text{-}129, 7\text{-}110,\\ 40\text{-}68, 54\text{-}123,\\ 61\text{-}66, 64\text{-}67\\ 65\text{-}66, 68\text{-}73,\\ 68\text{-}173, 174\text{-}198,\\ 185\text{-}186 \}\end{array}$& SSRP+BFS& $\begin{array}{c}\{ 3\text{-}129, 7\text{-}110,\\ 40\text{-}68, 54\text{-}123,\\ 57\text{-}180, 57\text{-}190,\\ 66\text{-}190, 67\text{-}190,\\ 68\text{-}73, 68\text{-}173,\\ 168\text{-}187, 172\text{-}187,\\ 174\text{-}198, 184\text{-}185 \}\end{array}$  &  $77.187$ & $77.187 $ & $77.187 $ &  $\begin{array}{c}\{ 19.76,\\ 6.11,\\ 205.98,\\ 76.9\}\end{array}$ &  $\begin{array}{c}\{114.76,\\ 6.11,\\ 110.98,\\ 76.9\}\end{array}$& 1908.2\\ \hline

IEEE 300  & 4  & 3 &$\begin{array}{c}\{57\text{-}66, 64\text{-}67,\\ 66\text{-}190, 68\text{-}173,\\ 109\text{-}110,  109\text{-}129,\\ 122\text{-}123, 174\text{-}191,\\ 174\text{-}198, 184\text{-}185,\\ 185\text{-}187 \}\end{array}$& \cite{kyriacou2017controlled}& $\begin{array}{c}\{ 7\text{-}110, 57\text{-}66,\\ 66\text{-}190, 67\text{-}190,\\ 68\text{-}173, 109\text{-}129\\ 122\text{-}123, 168\text{-}187,\\ 172\text{-}187, 174\text{-}191,\\ 174\text{-}198, 184\text{-}185 \}\end{array}$  &  $77.187$ & $77.187 $ & $77.187 $ &  $\begin{array}{c}\{ 145.98,\\ 79.76,\\ 6.11\\ 76.9\}\end{array}$ &  $\begin{array}{c}\{110.98,\\ 114.76,\\ 6.11,\\76.9\}\end{array}$& 1908.2\\
\bottomrule
\end{tabular}
}
\caption{Results after applying Algorithms \ref{alg:Migr_alg_a} and \ref{alg:Migr_alg_b} to the IEEE 300 test cases, considering different initial partitions $\Pi(0)$. 
Power values are reported in MW. Note that bound \eqref{eq:fin_bound} is computed for $k=K$.}
\label{Tab:migration_results_300}
\end{center}
\end{table*}

\section{Proof of Convergence}
\label{sec:proof}

To prove Lemma \ref{lemma1} and Theorem \ref{thm:main} we first need to define the stack vector $\B{P}(k)\coloneqq[P_1(k) \ \cdots \ P_{n_\mu}(k)]\T$ and $\B{P}^* \coloneqq p^* \B{1}$, and then give the following Lemma.

\begin{proof}
From \eqref{eq:cost}, 
we have that 
\begin{equation}\label{eq:sum_abs_P}
         J(k) = \frac{1}{n_\mu} \left(\sum_{l:  P_l(k)>0} P_l(k) - \sum_{l: P_l(k)\leq 0} P_l(k)\right).
\end{equation}
Moreover, as 
$$\sum_{l:  P_l(k)>0} P_l(k) + \sum_{l: P_l(k)\leq 0} P_l(k) = \Ptot=n_{\mu}p^*,$$
we can recast \eqref{eq:sum_abs_P} as
\begin{equation*}
         J(k) = \frac{1}{n_\mu} \left(2\sum_{l:  P_l(k)>0} P_l(k) - n_{\mu}p^*\right). \label{eq:sum_abs_P_b}
\end{equation*}
Hence, as $J^*=|p^*|$ [from \eqref{eq:cost_bound}], we obtain
\begin{equation}\label{eq:J-J^*V2}
        J(k) - J^* = \frac{2}{n_{\mu}}\sum_{l:  P_l(k)>0} P_l(k) - (p^* +|p^*|).
\end{equation}
Without loss of generality, let us relabel the islands so that $P_1(k)\leq P_2(k)\leq \dots \leq P_{n_{\mu}}(k)$. 
Then, as the graph $\C{G}$ (defined in~§ \ref{sec:section_1}) and all the islands remain connected for all $k$, at each step also the graph $\C{G}^{\Pi(k)}$ (defined in~§ \ref{sec:section_1}) will be connected and thus \eqref{eq:neigh_bound} implies that
    \begin{equation}\label{eq:bound_noabs}
        P_{l+1}(k)\leq P_{l}(k) + \max_{i \in \C{V}}|p_i|, 
        \quad \forall l \in \{1, \dots, n_\mu - 1\}.
    \end{equation}
    Note that, from \eqref{eq:P_tot}, $\sum_{l=1}^{n_{\mu}}P_l(k) = \Ptot = n_{\mu}p^*$, and hence from \eqref{eq:bound_noabs} we obtain
    \begin{equation}\label{eq:P_l_worst_v2}
        P_l(k) \leq p^* + \bar{p} \left(l - \frac{n_\mu + 1}{2} \right),
        \quad \forall l \in \{1, \dots, n_\mu\},
    \end{equation}
    with $\bar{p} \coloneqq \max_{i \in \C{V}}|p_i|$. 
    From \eqref{eq:J-J^*V2}, $J(k)-J^*$ is maximized (worst case) when \eqref{eq:P_l_worst_v2} is an equality. 
    In such a case, to compute $J(k)-J^*$ by leveraging \eqref{eq:J-J^*V2}, we must first find 
    \begin{equation}\label{eq:l_star_def}
        l^*: \quad 
        P_l(k) \geq 0, \ \forall l \in \{ l^*, \dots, n_\mu\}.
    \end{equation}
    Hence, to find $l^*$ we must find the smallest integer $l$ such that 
    \begin{equation}\label{eq:ineq_l_star}
           p^* + \bar{p} \left(l - \frac{n_\mu + 1}{2} \right)\geq 0,
    \end{equation}
    yielding \eqref{eq:l_star}.
    Then, from \eqref{eq:l_star_def}, \eqref{eq:P_l_worst_v2}, and \eqref{eq:J-J^*V2}, we obtain \eqref{eq:fin_bound} and the Lemma is proved.
\end{proof} 
Now, let us exploit Lemma \ref{lemma1} to prove Theorem \ref{thm:main}.
    
    
    \begin{proof}[Proof of Theorem \ref{thm:main}]

   Consider a triplet $(l,m,i)$ fulfilling \eqref{eq:hypothesis}, and
    \begin{equation}\label{eq:comp_cond}
       |P_m(k) - P_l(k)| > |p_i|;
    \end{equation}   
   we start by showing that, when assuming \eqref{eq:hypothesis}, \eqref{eq:comp_cond} is equivalent to \eqref{eq:migration_rule}, i.e., a migration of node $i$ from island $\C{M}_m$ to $\C{M}_l$ will occur.
   
   Firstly, we show that \eqref{eq:migration_rule} implies \eqref{eq:comp_cond}. 
   When $P_m(k) < P_l(k)$, we have $p_i<0$ from \eqref{eq:hypothesis_2_1}, and from \eqref{eq:migration_rule} we have that
    \begin{equation}\label{eq:expand_mig_rule_1}
         P_m(k) < P_l(k+1).
    \end{equation}
  Differently, when $P_l(k) < P_m(k)$, we have $p_i>0$ from \eqref{eq:hypothesis_2_1}, and from \eqref{eq:migration_rule} we have that
    \begin{equation}\label{eq:expand_mig_rule_2}
        P_l(k) <  P_m(k+1).
    \end{equation}
    From \eqref{eq:expand_mig_rule_1} and \eqref{eq:expand_mig_rule_2}, recalling \eqref{eq:migration_rule_c_1} and \eqref{eq:migration_rule_c_2}, we have
    \begin{equation}\label{eq:cond_final}
        \begin{dcases}
          P_m(k)-P_l(k) <  p_i, & \text{if } p_i < 0,  \\  
          P_m(k)-P_l(k) >  p_i,  & \text{if } p_i > 0.
        \end{dcases}
    \end{equation}
    As \eqref{eq:cond_final} implies \eqref{eq:comp_cond}, we have proved that \eqref{eq:migration_rule} implies \eqref{eq:comp_cond}.
    
    Now, let us prove that \eqref{eq:comp_cond} implies \eqref{eq:migration_rule}.
    To do so, note that \eqref{eq:comp_cond} is equivalent to 
   \begin{equation}\label{eq:cond_final_1}
   \begin{dcases}
         P_l(k) > P_m(k)+ |p_i| , & \text{if } P_l(k) > P_m(k),\\
         P_m(k) > P_l(k) + |p_i|, & \text{if } P_l(k) < P_m(k).
   \end{dcases}         
   \end{equation}
   Moreover, exploiting \eqref{eq:hypothesis_2_1} and  recalling \eqref{eq:migration_rule_c_1} and \eqref{eq:migration_rule_c_2}, \eqref{eq:cond_final_1} can be recast as 
\begin{equation}\label{eq:relationMR_1}
    \begin{dcases}
        P_m(k) < P_l(k) + p_i = P_l(k+1) , & \text{if } P_l(k) > P_m(k),\\
        P_l(k) > P_m(k) -  p_i = P_m(k+1), & \text{if } P_l(k) < P_m(k).
    \end{dcases}
   \end{equation}
   It is straightforward to see that \eqref{eq:relationMR_1} immediately leads to \eqref{eq:migration_rule}. 
   Therefore, we have proved that (when \eqref{eq:hypothesis} holds) $\eqref{eq:comp_cond} \Leftrightarrow \eqref{eq:migration_rule}$.%

   As \eqref{eq:migration_rule} is equivalent to \eqref{eq:comp_cond} and \eqref{eq:hypothesis}, then if at some step, say $K$, no triplet $( l,m,i )$ existed fulfilling \eqref{eq:comp_cond}, the migration process would stop and, as the network $\C{G}$ is connected and so is the graph $\C{G}^{\Pi(K)}$ at that step, we would have 
   \begin{equation}\label{eq:temp_b}
        |P_l(K)-P_m(K)|\leq\max_{i \in \C{V}}|p_i| \quad  \forall l,m:\C{V}_m(K)\cap\C{N}(\C{V}_l(K)) \neq \varnothing.
    \end{equation}
   As from Lemma \ref{lemma1}, \eqref{eq:temp_b} implies that the bound \eqref{eq:fin_bound} holds, to prove our thesis we are left with showing that a stopping time instant $K$ exists. Firstly, note that such a step $K$  exists if \eqref{eq:migration_rule} fulfills    
   \begin{equation}\label{eq:contraction}
             \left\lVert \B{P}(k+1) - \B{P}^* \right\rVert_2 \leq \alpha \left\lVert \B{P}(k) - \B{P}^* \right\rVert_2 \  \forall k \in \{0, ... , K-1\}
    \end{equation}
    for some positive scalar $\alpha<1$ as if  \eqref{eq:contraction} were satisfied, then our migration rule would be a contraction mapping. In such case, from the Banach-Caccioppoli theorem \cite{kirk2002handbook}, there would be no limit cycles in the sequence $\lbrace \B{P}(k)\rbrace$ and thus also in $\lbrace \Pi (k) \rbrace$. Hence, as the number of possible partitions is finite, so would be the sequence $\lbrace \B{P}(k)\rbrace$ and thus, to complete our proof, we need to show that \eqref{eq:migration_rule} implies \eqref{eq:contraction}. As we have enforced that only one migration occurs at each step $k$, then $\B{P}(k+1)$ only differs from $\B{P}(k)$ for the $l$-th and $m$-th entries. Hence, proving \eqref{eq:contraction} only requires showing that
    \begin{equation}\label{eq:contraction_2b}
        \begin{split}
            (P_l(k+1)-p^*)^2+&(P_m(k+1)-p^*)^2 \\
            & < (P_l(k) - p^*)^2 + ( P_m(k) - p^* )^2
        \end{split}
    \end{equation}
    for all $k \in \{0, ... , K-1\}$.
    After a few algebraic simplifications, \eqref{eq:contraction_2b} can be rewritten as
    \begin{equation}\label{eq:discriminant}
       p_i (P_l(k) - P_m(k) + p_i) < 0 \quad  k \in \{0, ... , K-1\},
   \end{equation}
   which is trivially fulfilled by any triplet $(l,m,i)$ fulfilling \eqref{eq:hypothesis} and \eqref{eq:comp_cond}, yielding that  \eqref{eq:hypothesis} and \eqref{eq:comp_cond} imply \eqref{eq:contraction}. In turn, as \eqref{eq:hypothesis} and  \eqref{eq:comp_cond} imply \eqref{eq:migration_rule}, the existence of $K$ and thus our thesis remains proved.
\end{proof}

\section{Conclusions}

We introduced a power network islanding algorithm that solves the Intentional Controlled Islanding problem in a distributed manner. 
Our strategy allows the network nodes to self-organise so as to minimize the average absolute power imbalance among islands.
To allow the nodes to make informed decisions, we devised a consensus-based estimator which is instrumental to the migration process, as it allows nodes to estimate the power imbalances of neighboring islands in a distributed manner.
We demonstrated analytically that our algorithm converges in finite time to a partition whose average absolute power imbalance is in a given neighborhood of the optimal one. 
We tested the strategy on two benchmark power networks, the IEEE 118 and 300 bus systems, after the disconnection of one of their transmission lines showing the effectiveness of the proposed approach.

\bibliographystyle{IEEEtran}
\bibliography{bibliography}%

%

\end{document}